\documentclass[submission,copyright,creativecommons]{eptcs}
\providecommand{\event}{FROM 2022} 

\usepackage{iftex}

\ifpdf
  \usepackage{underscore}         
  \usepackage[T1]{fontenc}        
\else
  \usepackage{breakurl}           
\fi

\usepackage{graphicx}
\usepackage[disable]{todonotes}
\usepackage{makecell}
\usepackage{proof}
\usepackage{arydshln}
\usepackage{thm-restate}
\usepackage{amsmath}
\usepackage{amssymb}
\usepackage{xcolor}
\usepackage{xspace}
\usepackage{wrapfig, framed}
\usepackage{mdframed}
\usepackage{multicol}
\usepackage{mathpartir}
\usepackage{listings}
\usepackage{stmaryrd}
\usepackage[inline]{enumitem}
\usepackage{hyperref}
\usepackage{amsthm}
\newcommand{\ML}{\ensuremath{\textsc{ML}}\xspace}

\newcommand{\var}{\mathit{var}}
\newcommand{\dom}{\mathit{dom}}

\newcommand{\eqbydef}{\triangleq}
\newcommand{\dfness}[1]{\lceil#1\rceil}
\newcommand{\tlness}[1]{\lfloor#1\rfloor}

\newcommand{\mdot}{\boldsymbol{\cdot}}

\newcommand{\evar}{\mathit{EVar}}
\newcommand{\svar}{\mathit{SVar}}

\newcommand{\lgg}{\textit{lgg}\xspace}

\newcommand{\substexists}{\textsc{$\exists$-{Subst}}\xspace}

\newcommand{\existsgen}{\textsc{$\exists$-{Gen}}\xspace}

\newcommand{\eqvcontext}{\leftrightarrow_{\mathit{context}}}
\newcommand{\eqvtranz}{\leftrightarrow_{\mathit{tranz}}}
\newcommand{\econtext}{\textsc{$\exists$-Ctx}\xspace}
\newcommand{\escope}{\textsc{$\exists$-Scope}\xspace}
\newcommand{\ecollapse}{\textsc{$\exists$-Collapse}\xspace}
\newcommand{\adec}{\textsc{Dec}\xspace}

\newcommand{\orgen}{$\vee_\mathit{gen}$\xspace}
\newcommand{\decaunif}{$\rightsquigarrow_\mathit{step}$\xspace}

\newcommand{\aaa}[2][]{\todo[author=Andrei,color=green!10!green,#1]{#2}}
\newcommand{\dl}[2][]{\todo[author=Dorel,color=red!10!white,#1]{#2}}

\definecolor{mygray}{gray}{0.4}
\newcommand{\change}[1]{\textcolor{mygray}{#1}}

\newcommand{\Smbs}{{\textsf{Symbol}:}}
\newcommand{\Axms}{{\textsf{Axiom}:}}
\newcommand{\Sugar}{{\textsf{Notation}:}}
\newcommand{\inh}[1]{[\![{#1}]\!]}
\newcommand{\cln}{{:}}
\newcommand{\ld}{\mathord{.\,}}
\newcommand{\limplies}{\rightarrow}
\newcommand{\imp}{\limplies}
\newcommand{\Impts}{{\textsf{Import}:}}
\newcommand{\axname}[1]{\textnormal{(\textsc{#1})}}

\newcommand{\snext}{{\bullet}}

\newtheorem{example}{Example}
\newtheorem{theorem}{Theorem}

\newtheorem{remark}{Remark}
\newtheorem{definition}{Definition}

\lstdefinelanguage{AML}
{morekeywords={spec, endspec, model, endmodel, of},
sensitive=false,
morecomment=[l][\textit]{//},
morecomment=[s]{/*}{*/},
keywordstyle=\bfseries\sffamily,
backgroundcolor=\color{white},
}

\title{Proof-Carrying Parameters in Certified Symbolic Execution:\\ The Case Study of Antiunification}
\author{Andrei Arusoaie \qquad\qquad Dorel Lucanu
\institute{Faculty of Computer Science}
\institute{\emph{Alexandru Ioan Cuza}, University of Iași, România}
\email{\{arusoaie.andrei,dlucanu\}@info.uaic.ro}
}
\def\titlerunning{Antiunification in matching logic}
\def\authorrunning{A. Arusoaie\& D. Lucanu}
\begin{document}
\maketitle

\begin{abstract}
  Symbolic execution  uses various algorithms (matching, (anti)unification), whose executions are parameters for proof object generation.  This paper proposes a generic method for generating proof objects for such parameters. We present in detail how our method works for the case of antiunification. The approach  is accompanied by an implementation prototype, including a proof object generator and a proof object checker.   In order to investigate the size of the proof objects, we generate and check proof objects for inputs inspired from the K definitions of C and Java.  
\end{abstract}

\section{Introduction}

K (\url{https://kframework.org}) is a well established framework for programming languages, which brings a different perspective of what such a framework should be.
K provides means to give formal definitions for programming languages and aims to automatically derive a series of practical tools for those languages: a parser, an interpreter, a debugger, a symbolic execution tool, a deductive verifier, a model-checker, and others.
A formal semantics of a language defined in K consists of syntax declarations, a language configuration, and a set of rewriting rules. The configuration is a constructor term which holds the semantical information needed to execute programs (e.g., the code, environment, stack, program counter, etc).
The rewriting rules are pairs $\varphi \Rightarrow \varphi'$ of program configurations with variables which specify how program configurations transit to other program configurations.
An example of a tool automatically generated by K is the interpreter, which works as follows: the user provides a concrete configuration (which includes the program and an initial state of that program) and K applies rewriting rules as much as possible to this configuration. Another tool is the K prover, which uses symbolic execution to prove reachability properties. 

The theoretical foundation of K is Matching Logic~\cite{rosu-2017-lmcs,chen-lucanu-rosu-2021-jlamp} (hereafter shorthanded as \ML), a logical framework where the formal definitions of program languages~\cite{ellison-rosu-2012-popl,DBLP:conf/pldi/HathhornER15,park-stefanescu-rosu-2015-pldi} and program reasoning~\cite{stefanescu-park-yuwen-li-rosu-2016-oopsla,stefanescu-ciobaca-mereuta-moore-serbanuta-rosu-2014-rta,DBLP:conf/wrla/RusuA16,DBLP:conf/birthday/LucanuRAN15,arusoaie:hal-01627517} can be done in a uniform way. 
\ML formulas are called \emph{patterns} and they are used to uniformly specify syntax and semantics of programming languages, and the properties of program executions.  
\ML has a \emph{pattern matching semantics}: a pattern is interpreted as the set of elements that \emph{match} it. 
For example, if a pattern $t$ encodes a symbolic program configuration (with variables), then $t$ is interpreted as the set of concrete program configurations that match it. 

\ML has a minimal, but expressive, syntax.
For example, if one wants to specify configurations $t$ that satisfy a first-order constraint $\phi$, then $t \land \phi$ is the pattern that captures this intent.
Not only constraints can be attached to patterns. 
Conjunctions $t_1 \land t_2$ (and disjunctions $t_1 \lor t_2$) are interpreted as the intersection (and union, respectively) of the elements that match $t_1$ and $t_2$.
Moreover, it is easy to explain K rules $\varphi \Rightarrow \varphi'$ as implications of patterns $\varphi\imp \snext\varphi'$: the program configurations that match $\varphi$ can transit in one step to program configurations that match $\varphi'$, where $\snext$ is a special symbol used to specify one step transitions.

\ML is equipped with a sound proof system which can derive sequents of the form $\Gamma \vdash \varphi$, where $\varphi$ is a pattern and $\Gamma$ is an \emph{\ML theory} (i.e., a set of axiom patterns). 
The \ML proof system is the key ingredient of another tool that K aims to generate: a deductive verifier.\\

\textbf{Motivation.} A fair question that needs to be posed is \emph{how can we trust the proofs produced by the deductive verifier generated by K?} Given the size of the K codebase (about half a million lines of code~\cite{chen-lin-trinh-rosu-2021-cav}) and its dynamics (new code committed every week), the formal verification of the implementation of K is out of question. The solution here is to do what other formal verification tools 
do: instrument K so that its automatically generated tools produce \emph{proof objects} that can be independently checked by a trusted kernel. In our context, proof objects are just proofs that use the \ML proof system. 

It turns out that all these tools that K aims to generate share several components. For example, \emph{matching} algorithms are useful for concrete execution (interpreter), while \emph{unification} and \emph{antiunification} algorithms are needed for symbolic execution and program verification. 
Therefore, we can have a uniform approach: if we can find proof object generation mechanisms for each component, then we can simply instantiate those mechanisms whenever needed. 

Symbolic execution is a key component 
in program verification and it has been used in K as well (e.g., ~\cite{arusoaie-2014,JSC2016,stefanescu-park-yuwen-li-rosu-2016-oopsla}). 
Generating proof objects for symbolic execution is difficult because the  parameters of an execution step must carry more proof information than in the concrete executions case. First, instead of matching, proof parameters must include unification information. Second, path conditions need to be carried along the execution. 

In \ML, there is a natural way to deal with symbolic execution. \ML patterns $\varphi$ have a \emph{normal form} $t \land \phi$, where $t$ is a term pattern and $\phi$ is a predicate pattern, expressing a constraint on variables in $t$. 
In particular, $t$ can be the program configuration and $\phi$ the path condition.
Patterns $t \land \phi$ are evaluated to the set of values that \emph{match} $t$ and satisfy $\phi$. 
To compute the symbolic successors of a pattern, say $t \land \phi$, with respect to a rule, say $t_1 \land \phi_1  \Rightarrow t_2 \land \phi_2$, we need to unify the patterns $t \land \phi$ and $t_1 \land \phi_1$. 
Because unification can be expressed as a conjunction in \ML~\cite{fm,rosu-2017-lmcs}, we can say that only the states matched by $(t \land \phi) \land (t_1 \land \phi_1) \equiv (t \land t_1) \land (\phi \land \phi_1)$ transit to states matched by $t_2 \land \phi_2$. 
Expressing unification as a conjunction $(t \land t_1)$ is a nice feature of \ML, but, in practice, unification algorithms are still needed to compute the general unifying substitution since it is used in symbolic successor patterns. The symbolic successors are obtained by applying the unifying substitution to the right-hand side of a rule (e.g., $t_2 \land \phi_2$) and adding the substitution (as an \ML formula) to the path condition. 
Also, unification algorithms are being used to normalise conjunctions of the form $t \land t_1$, so that they consist of only one term and a constraint, $t'\land \phi'$. 
Therefore, unification algorithms are parameters of the symbolic execution steps and they must be used to generate the corresponding proof objects.

It is often the case when more than one rule can be applied during symbolic execution. For instance, if an additional rule $t_1' \land \phi_1'  \Rightarrow (t_2' \land \phi_2')$ can  be  applied to $t \land \phi$, then the set of target states must match $t_2 \land \phi_2$ or $t_2' \land \phi_2'$. This set of states is matched by the disjunction $(t_2 \land \phi_2) \vee (t_2' \land \phi_2')$. For the case when $\phi_2 \land \phi'_2$ holds, the disjunction reduces to $t_2 \lor t'_2$, which is not a normal form but it can be normalised using antiunification.\\

\textbf{Related work.}
The literature on (anti)unification is vast. Due to the space limit, 
we recall here the closest related work which addresses proof object generation for concrete and symbolic executions, to understand the context of our work.
In~\cite{chen-lin-trinh-rosu-2021-cav}, the authors propose a method to generate proof objects for program executions
$\varphi_\mathit{init} \Rightarrow \varphi_\mathit{final}$, 
where $\varphi_\mathit{init}$ is the formula that specifies the initial state of the execution, $\varphi_\mathit{final}$ specifies the final state, and ``$\Rightarrow$'' states the rewriting/reachability relation between states.  The correctness of an execution,\\

\centerline{$
\Gamma \vdash \varphi_\mathit{init} \Rightarrow \varphi_\mathit{final},
$}

\noindent
 is witnessed by a formal proof, which uses the \ML proof system. The K interpreter computes the parameters (e.g., execution traces, matching info) needed to generate the proof object.

In~\cite{rosu-2017-lmcs}, the author shows that unification in \ML can be represented as a conjunction of \ML patterns.
A first step into generating proof objects for unification was done in~\cite{fm}. 
We proposed a method to normalise conjunctions of patterns $t_1 \land t_2$. 
The K implementation works with patterns in normal form $t \land \phi$, which are more efficient: matching/unification algorithms are executed only once on normalised patterns, rather than multiple times on patterns having multiple structural components (e.g., $t_1 \land t_2$). 
In~\cite{fm}, we use the syntactic unification algorithm~\cite{martelli} to (1) find an equivalent normal form $t \land \phi$ for conjunctions $t_1 \land t_2$, and (2) to generate proof objects for the equivalence between $t \land \phi$ and $t_1 \land t_2$. 
The unification algorithm provides the needed parameters (e.g., unifying substitutions) for proof generation.\\
%
 
 \textbf{Contributions.}
 A lesson that we learned from~\cite{fm} and~\cite{chen-lin-trinh-rosu-2021-cav} is that the algorithms implemented in various components of K can be used to compute the parameters needed to generate proof objects. 
 In this paper we address the problem of \emph{generating proof objects for antiunification}, which is used, e.g., in symbolic execution and verification. 
 In \ML, the \emph{least general generalisation} of two term patterns $t_1$ and $t_2$ is given by their disjunction $t_1 \lor t_2$. 
 We use Plotkin's antiunification algorihm~\cite{Plo:72,Plotkin70} to find normal forms $t \land \phi$ for disjunctions $t_1 \lor t_2$, and to generate proof objects for the equivalences between $t \land \phi$ and $t_1 \lor t_2$. 
 The execution of the antiunification algorithm provides the parameters (intermediate generalisations and substitutions computed at each step) to generate the proof objects. Our~contributions~are:
 \begin{enumerate}
   \item 
We express Plotkin's antiunification algorithm in \ML terms and we show that its steps produce equivalent patterns (Lemma~\ref{lem:stepaunif} and Theorem~\ref{th:antiunif}).
   \item 
We propose a proof object generation mechanism for the equivalences computed by the algorithms used in symbolic execution and verification, and we show how it works in the case of antiunification (Section~\ref{sec:aunifgen}).
   \item We provide a prototype implementation of our proof object generation mechanism and a proof checker (Section~\ref{sec:prototype});
   \item We test our prototype on interesting examples, including inputs inspired from the K definitions of C~\cite{ellison-rosu-2012-popl,DBLP:conf/pldi/HathhornER15} and Java~\cite{DBLP:conf/popl/BogdanasR15}.
 \end{enumerate}
Indeed, the most challenging part of this work is the proof object generation mechanism. 
The main difficulty was to find the right proof object schema generation that precisely captures one step of the antiunification algorithm. 
Another tricky part was to design the proof object schema so that the proofs generated for each step can be easily composed. 
The size of the resulted proofs depends on the number of steps performed by the antiunification algorithm.\\

\textbf{Paper organisation.}
Section~\ref{sec:aml} presents \ML, its proof system and the \ML theory for many-sorted term algebras.
In Section~\ref{sec:aunif} we present antiunification in a \ML setting, and we prove that Plotkin's antiunification can be safely used to normalise disjunctions of term patterns.
Our proof object generation methodology is presented in Section~\ref{sec:aunifgen}. 
The prototype implementation is described in Section~\ref{sec:prototype} and we conclude in Section~\ref{sec:conclusions}.

\section{Matching Logic}
\label{sec:aml}

Matching logic (\ML)~\cite{rosu-2017-lmcs,CR19,chen-lucanu-rosu-2021-jlamp} started as a logic over a particular case of constrained terms~\cite{DBLP:conf/lics/RosuSCM13,stefanescu-ciobaca-mereuta-moore-serbanuta-rosu-2014-rta,arusoaie:hal-01627517,DBLP:conf/birthday/LucanuRAN15}, but now it is developed as a full logical framework. 
We recall from~\cite{chen-lucanu-rosu-2021-jlamp} the definitions and notions that we use in this paper. 

A matching logic \emph{signature} is a triple $(\evar, \svar, \Sigma)$, where $\evar$ is a set of \emph{element variables} $x, y, \ldots$, $\svar$ is a set of \emph{set variables} $X,Y,\ldots$, and $\Sigma$ is a set of \emph{constant symbols} (or \emph{constants}). 
The set \textsc{Pattern} of \emph{$\Sigma$-patterns} is generated by the grammar below, where $x \in \evar$, $X \in \svar$, and $\sigma \in \Sigma$:
\\[1ex]
\centerline{$
\varphi ::= x \mid X \mid \sigma \mid \varphi_1~\varphi_2 \mid \bot \mid \dfness{\varphi} \mid \varphi_1 \rightarrow \varphi_2 \mid \exists x . \varphi \mid \mu X . \varphi \textit{ if $\varphi$ is positive in $X$}
$}\\[1ex]
A pattern $\varphi$ is \emph{positive} in $X$ if all free occurrences of $X$ in $\varphi$ are under an even number of negations.
The patterns below are derived constructs:\\[1ex]
\centerline{$
\begin{aligned}
\top &\equiv \lnot \bot & \tlness{\varphi} &\equiv \lnot \dfness{\lnot \varphi} & \varphi_1 \lor \varphi_2 &\equiv \lnot \varphi_1 \rightarrow \varphi_2  \\
\lnot \varphi &\equiv \varphi \rightarrow \bot & \varphi_1 = \varphi_2 &\equiv \tlness{\varphi_1 \leftrightarrow \varphi_2} & \varphi_1 \land \varphi_2 &\equiv \lnot (\lnot \varphi_1 \lor \lnot \varphi_2) & \\
\forall x. \varphi &\equiv \lnot\exists x . \lnot \varphi & \varphi_1 \neq \varphi_2 &\equiv \lnot (\varphi_1 = \varphi_2) &  \varphi_1 \leftrightarrow \varphi_2 &\equiv (\varphi_1 \rightarrow \varphi_2) \land (\varphi_2 \rightarrow \varphi_1)
\end{aligned}
$\vspace{-1ex}}
\begin{example}
\label{ex:signature}
Let $\Sigma = \{ \mathit{zero}, \mathit{succ}, \mathit{nil}, \mathit{cons} \}$ be an \ML signature.
Then $x$, $\mathit{zero}$, $\mathit{succ}\,\mathit{zero}$, $\mathit{succ}\,x$, 
$\exists x. \mathit{zero} = x$, $\mu X. \mathit{zero} \lor (\mathit{succ}\, X)$
are examples of \ML patterns\footnote{Of course, $\mathit{succ}\,\mathit{nil}$ or $\mathit{cons}\,\mathit{nil}\,\mathit{zero}$ are also \ML patterns but these can be handled properly using sorts (see Section~\ref{sec:termalg}).}.

\vspace{-1ex}
\end{example}
%

\ML has a pattern matching semantics where patterns are interpreted on a given carrier set
, say $M$.
 Each pattern is interpreted as the set of elements that match it.
\emph{Element variables} $x$ are matched by a singleton set, while \emph{set variables} $X$ are matched by a subset of $M$. 
The pattern $\bot$ is matched by the empty set (and hence $\top$ by $M$).
The implication pattern $\varphi_1 \rightarrow \varphi_2$ is matched by the elements that do not match $\varphi_1$ or match $\varphi_2$.
A pattern $\exists x\ld\varphi$ is matched by the instances of $\varphi$ when $x$ ranges over $M$. In particular, $\exists x.x$ is matched by $M$. 
Note that $\exists$ binds only element variables. 
Symbols $\sigma$ (e.g. $\mathit{zero}$, $\mathit{succ}$) are interpreted as subsets $\sigma_M\subseteq M$, and, usually, the needed interpretation for them is obtained using axioms. For instance,  the pattern $\exists x.\mathit{zero}=x$ is matched by $M$ if $\mathit{zero}_M$ is a singleton, and by $\bot$ otherwise. 
This type of pattern is often used as axiom to restrict the interpretation of symbols to singletons.
The pattern $\varphi_1~\varphi_2$ is an \emph{application} and its interpretation is given by means of a function $M\times M\to \mathcal{P}(M)$, which is pointwise extended to a function $\mathcal{P}(M)\times \mathcal{P}(M)\to \mathcal{P}(M)$. Applications are useful to build various structures or relations. 
For instance, $\forall x\ld \exists y\ld \mathit{succ}\,x=y$ says that $\mathit{succ}$ has a functional interpretation (recall that the element variable $y$ is matched by a singleton set).
Applications are left associative.
The pattern $\mu X\ld\varphi$ is matched by the least fixpoint of the functional defined by $\varphi$ when $X$ ranges $\mathcal{P}(M)$. An example is $\mu X. (\mathit{zero}\lor \mathit{succ}\,X)$, which is matched by the natural numbers $\mathbb{N}$ (up to a surjection), when both $\mathit{zero}$ and $\mathit{succ}$ have a functional interpretation (as above).
Note that $\mu$ binds only set variables in \emph{positive} patterns.
 %
%
%
The pattern $\dfness{\varphi}$ is called \emph{definedness}\footnote{For convenience, we introduce it directly in the syntax of patterns but it can be axiomatised as in~\cite{rosu-2017-lmcs}.} and it is matched by $M$ if $\varphi$ is matched at least by one element, and by $\emptyset$ otherwise. Such patterns are called \emph{predicate patterns}. 

The syntax priorities of the \ML constructs is given by this ordered list:\\[0.5ex]
\centerline{$
\lnot\_, \dfness{\_}, \tlness{\_}, \_=\_,  \_\land\_, \_\lor\_, \_\rightarrow\_, \_\leftrightarrow\_, \exists\_.\_, \forall\_.\_, \mu\_.\_,
$}\\[0.5ex]
where $\lnot$ has the highest priority and $\mu\_.\_$ has the lowest priority.
By convention, the scope of the binders extends as much as possible to the right, and parentheses can be used to restrict the scope of the binders.
We often write $\varphi[\psi/x]$ and $\varphi[\psi/X]$ to denote the pattern obtained by substituting all free occurrences of $x$ and $X$, respectively, in $\varphi$ for $\psi$. In order to avoid variable capturing, we consider that $\alpha$-renaming happens implicitly.

\emph{The \ML proof system}~\cite{chen-lucanu-rosu-2021-jlamp}
is shown in Figure~\ref{fig:proofsystem}. 
It contains four categories of rules: propositional tautologies, frame reasoning over application contexts, standard fixpoint reasoning, and two rules needed for completeness. An \emph{application context} $C$ is a pattern with a distinguished placeholder variable $\square$ s.t. the path from the root of $C$ to $\square$ has only applications. $C[\varphi/\square]$ is a shorthand for  $C[\varphi]$ and $\mathit{free}(\varphi)$ denotes the set of free variables in $\varphi$.

\begin{figure}[t]
\selectfont
\centering
\renewcommand{\arraystretch}{1.15}
\begin{tabular}{|lrl|}
\hline
\multicolumn{3}{|c|}{\textbf{Hilbert-style proof system}}\\ 
\hline
\textsc{Propositional} && $\varphi$,  if $\varphi$ is a propositional tautology over patterns\\
\textsc{Modus Ponens} && $\infer{\varphi_2}{\varphi_1 & \varphi_1 \rightarrow \varphi_2}$ \\
\textsc{$\exists$-Quantifier} && $\varphi[y/x]\rightarrow \exists x . \varphi$ \\
\textsc{$\exists$-Generalisation} && 
\begin{minipage}{.2\textwidth}$\infer{(\exists x . \varphi_1) \rightarrow \varphi_2}{\varphi_1 \rightarrow \varphi_2}$\end{minipage} if $x \not\in \mathit{free}(\varphi_2)$ \\
\hline
\textsc{Propagation$_\bot$} && $C[\bot] \rightarrow \bot$ \\
\textsc{Propagation$_\lor$} && $C[\varphi_1 \lor \varphi_2] \rightarrow C[\varphi_1] \lor C[\varphi_2]$ \\
\textsc{Propagation$_\exists$} && $C[\exists x . \varphi] \rightarrow \exists x . C[\varphi] $ ~~~ if $x \not\in \mathit{free}(C)$\\
\textsc{Framing} && $\infer{C[\varphi_1] \rightarrow C[\varphi_2]}{\varphi_1 \rightarrow \varphi_2}$\\
\hline
\textsc{Set Variable Substitution} && $\infer{\varphi[\psi/X]}{\varphi}$\\
\textsc{Pre-Fixpoint} && $\varphi[\mu X. \varphi / X] \rightarrow \mu X . \varphi$\\
\textsc{Knaster-Tarski} && $\infer{\mu X . \varphi \rightarrow \psi}{\varphi[\psi/X] \rightarrow \psi}$\\
\hline
\textsc{Existence} && $\exists x . x$\\
\textsc{Singleton} && $\lnot (C_1[x \land \varphi] \land C_2[x \land \lnot \varphi])$\\
\hline
%
%
%
\end{tabular}
\caption{The Hilbert-style \ML proof system.}
\label{fig:proofsystem}
\end{figure}
%


\subsection{\ML Specification of the Term Algebra} 
\label{sec:termalg}

\newcommand{\Sort}{\mathit{Sort}}
\newcommand{\Unit}{\mathit{Unit}}
\newcommand{\unit}{\mathit{unit}}

A complete \ML axiomatization of the many-sorted term algebra is given in~\cite{chen-lucanu-rosu-2020-tr} and we briefly recall it in Figure~\ref{spec:msa}. 
The specification $\mathsf{SORTS}$ introduces symbols for sorts and their inhabitant sets, and some usual notations for them.
The specification $\mathsf{MSA}$ includes the axioms corresponding to a given algebraic signature.
Finally, $\mathsf{TERM(S, F)}$ includes the properties "no confusion" and "no junk" (inductive domains) 
that characterizes the (initial) term algebra.
 "No confusion" says that the function symbols $F$ are constructors:
 \begin{itemize}[topsep=0pt, partopsep=0pt, itemsep=0pt]
\item two different constructors will define different terms $\axname{NoConfusion I}$;
\item a constructor is injective (i.e., the same constructors with different arguments will define different terms), and this is captured by $\axname{NoConfusion II}$.
\end{itemize}
The "no junk" property says that all inhabitants of a sort are generated using the constructors $F$, and this is captured by the axiom \axname{Induct. Domain}.
For the sake of presentation, this axiom does not include the case of the mutual recursive sorts\footnote{See~\cite{chen-lucanu-rosu-2021-jlamp} for a complete definition.}.

\begin{figure}[h]
  \renewcommand\figurename{\textsf{\textbf{Specification}}}
  \small
  \begin{tabular}{|c|c|}
  \hline
  \begin{minipage}{0.44\textwidth}\vspace{-1\baselineskip}
  \begin{lstlisting}[mathescape]
spec $\mathsf{SORTS}$ 
 (*\Smbs*) $\mathit{inh},\Sort$
 (*\Sugar*)
  $\inh{s} \equiv \mathit{inh} \  s$ 
  $\forall x \cln s \ld \varphi \equiv \forall x \ld x \in \inh{s} \imp 
 \varphi$
  $\exists x \cln s \ld \varphi \equiv \exists x \ld x \in \inh{s} 
 \land 
 \varphi$
  $\varphi \cln s \equiv \exists z \cln s \ld \varphi = z$
  $\forall x_1,{\ldots},x_n \cln s \ld \varphi \equiv
  \forall x_1 \cln s  {\ldots} \forall x_n \cln s \ld \varphi$
  $\exists x_1,{\ldots},x_n \cln s \ld \varphi \equiv
  \exists x_1 \cln s {\ldots} \exists x_n \cln s \ld \varphi$
endspec
\end{lstlisting}\vspace{-1\baselineskip}
\end{minipage}
&
  \begin{minipage}{0.51\textwidth}\vspace{\baselineskip}\vspace{-0.75\baselineskip}
  \begin{lstlisting}[mathescape]
spec $\mathsf{MSA}(\ensuremath{S,F})$
 (*\Impts*) $\mathsf{SORTS}$
 (*\Smbs*) $s \in S$, $f \in F$
 (*\Sugar*)  
    $f(\varphi_1,\ldots,\varphi_n)\equiv f\,\varphi_1\,\ldots\,\varphi_n$
 (*\Axms*) 
 $\begin{array}{ll}
      \axname{Sort}& s\cln \Sort \textrm{~for~each~}s\in S\\
      \axname{NonEmpty}& \llbracket s \rrbracket \neq \bot\textrm{~for~each~}s\in S\\
      \axname{Function} & \forall x_1\cln s_1{\ldots}\forall x_n\cln s_n . (f\,x_1\ldots x_n) \cln s\\ 
        & \textrm{~for~each~}f \in F_{s_1\ldots s_n, s}
    \end{array}$
endspec
  \end{lstlisting}
  \end{minipage}
  \\
  \hline
  $\mathsf{SORTS}$
  &
  $\mathsf{MSA}$
  \\
\hline 
\end{tabular}
\centering
\\
  \begin{minipage}{1.0\textwidth}
  \begin{mdframed}[innerleftmargin=0.25em, innerrightmargin=0.25em, innertopmargin=.25em, innerbottommargin=.25em, skipabove=0.25cm, skipbelow=0.25cm]
  \begin{lstlisting}[mathescape]
spec $\mathit{TERM}(\ensuremath{S,F})$
 (*\Impts*) $\textsf{MSA}(\ensuremath{S,F})$ 
 (*\Axms*) 
 $\begin{array}{ll}
        \axname{NoConfusion I} & f\!\not=\!f'\!\rightarrow\!\forall x_1{:}s_1. .. \forall x_n{:}s_n.\forall x'_1{:}s'_1. .. \forall x'_m{:}s'_m. f\,x_1\,..\,x_n \neq f'\,x'_1\,..~x'_m\\
        \axname{NoConfusion II} & \forall x_1,x'_1{:}s_1. \ldots \forall x_n,x'_n{:}s_n. (f\,x_1\cdots x_n)=(f\,x'_1\cdots x'_n) \rightarrow\\
  & ~\hfill(x_1{=}x'_1)\land\cdots \land (x_n{=}x'_n)\\
       \axname{Induct. Domain} &\displaystyle \llbracket s \rrbracket = \mu X.\bigvee_{f:s_1\ldots s_n, s} f~Y_1\cdots Y_n,\textrm{~where~}  Y_i=\begin{cases} X, &\textrm{if~} s_i=s \\ \llbracket s_i\rrbracket, & \textrm{~otherwise}\end{cases}
   \end{array}$
endspec
  \end{lstlisting}\vspace{-0.25\baselineskip}
  \hrule\vspace{.1\baselineskip}
  \centerline{$\mathsf{TERM(S, F)}$}
  \end{mdframed}\vspace{-0.75\baselineskip}
\end{minipage}
  \caption{\ML specifications for sorts, many-sorted algebras, and term algebra}
  \label{spec:sort}
  \label{spec:term}
  \label{spec:msa}
  \end{figure}

\begin{theorem}[\cite{chen-lucanu-rosu-2020-tr}]
The specification $\mathsf{MSA}(S,F)$ captures the many-sorted $(S,F)$-algebras in the following sense:
\begin{itemize}
\item from each $\mathsf{MSA}(S,F)$-model $M$ we may extract an $(S,F)$-algebra $\alpha(M)$, and
\item for each $(S,F)$-algebra $A$ there is an $\mathsf{MSA}(S,F)$-model $M$ s.t.  $\alpha(M)=A$.
\end{itemize}
If $M$ is a $\mathit{TERM}(\ensuremath{S,F})$-model, then $\alpha(M)$ is the term $(S,F)$-algebra (up to isomorphism).
\end{theorem}

\begin{restatable}{proposition}{proprulestermalg}
\label{prop:rulestermalg}
The next patterns are semantical consequences of $\mathit{TERM(S, F)}$:\\[1ex]
\centerline{$
\begin{aligned}
\exists z. t \land (z = u) \leftrightarrow t[u/z] && \textrm{if~}z \not\in\var(u) \\
z=(f\, \overline{t}) \leftrightarrow \exists \overline{y}. z=(f\, \overline{y}) \wedge \overline{y} = \overline{t} && \textrm{if~}\overline{y} \not\in \var\big((f\, \overline{t})\big) \cup \{z\}
\end{aligned}$}\\[1ex]
\end{restatable}

\noindent
The equivalences in Proposition~\ref{prop:rulestermalg} are later used as macro rules for proof object generation
. The notation $(f\,\overline{t})$ means $(f\,t_1\,\ldots\,t_n)$. We also use
$\exists \overline{y}$ or $\exists \{y_1, \ldots, y_n\}$ instead of $\exists y_1.\ldots \exists y_n$. The equality $\overline{y} = \overline{t}$ is sugar syntax for $\bigwedge_{i=1}^n y_i = t_i$.

\section{Antiunification in \ML}
\label{sec:aunif}
\vspace{-1ex}

Antiunification is a process dual to unification~\cite{martelli} that computes a \emph{generalisation} $t$ of two input terms $t_1$ and $t_2$. 
A term $t$ is an \emph{antiunifier} of $t_1$ and $t_2$ if there are two substitutions $\sigma_1$ and $\sigma_2$ such that $t \sigma_1 = t_1$ and $t \sigma_2 = t_2$. There is at least one antiunifier for any $t_1$ and $t_2$: we can always choose a variable $x \not\in \mathit{var}(t_1, t_2)$ and substitutions $\sigma_1 = \{ x \mapsto t_1 \}$ and $\sigma_2 = \{ x \mapsto t_2 \}$ s.t. $x\sigma_1 = t_1$ and $x\sigma_2= t_2$. 


A term $t'$ is more \emph{general} than a term $t$ if there is a substitution $\sigma$ such that $t'\sigma = t$. Given $t_1$ and $t_2$, their \emph{least general} antiunifier $t$ satisfies: for any antiunifier $t'$ of $t_1$ and $t_2$ we have that $t$ is less general than $t'$ (a.k.a., \emph{least general generalisation}, shorthanded as \lgg).

\begin{remark}
In other words, $t$ is less {general} than $t'$ iff the set of ground instances of $t$ is included in that of $t'$. In terms of matching logic, this can be expressed by $\exists \var(t).t \subseteq \exists \var(t').t'$, where $\varphi_1\subseteq \varphi_2$ is defined as $\lfloor \varphi_1 \rightarrow \varphi_2\rfloor$.
\end{remark}

Now we present Plotkin's antiunification algorithm~\cite{Plotkin70} for computing the \lgg over \ML term patterns. First, we define \emph{antiunification problems}:

\begin{definition}
An \emph{antiunification problem} is a pair $\langle t, P \rangle$ consisting of a term pattern $t$ and a non-empty set $P$ of elements of the form $z \mapsto  u \sqcup v$, where $z$ is a variable, and $u$ and $v$ are term patterns.
\end{definition}

Plotkin's algorithm~\cite{Plotkin70} for computing the \lgg consists in applying a decomposition rule over antiunification problems as much as possible:\\ 

\noindent
$$\langle t, P \cup \{z\mapsto (f\,u_1\,\ldots\,u_n)\sqcup (f\,v_1\,\ldots\,v_n)\}\rangle \rightsquigarrow\langle t[(f\,z_1\,\ldots\,z_n)/z], P \cup \{z_1\mapsto u_1\sqcup v_1,\ldots, z_n\mapsto u_n\sqcup v_n\}\rangle,$$

\noindent
where $z_1,\ldots,z_n$ are fresh variables. If we want to compute the \lgg of $t_1$ and $t_2$, we build the initial antiunification problem $\langle z , \{ z \mapsto t_1  \sqcup t_2 \} \rangle$ with $z \not\in \var(t_1) \cup \var(t_2)$ and we apply Plotkin's rule repeatedly.
When this rule cannot be applied anymore, we say that the obtained antiunification problem $\langle t', P' \rangle$ is in \emph{solved form}. The obtained $t'$ is the \lgg of $t_1$ and $t_2$, while $P'$ defines the two substitutions $\sigma_1 =\{ z  \mapsto u \mid z \mapsto u \sqcup v \in P' \}$ and $\sigma_2  = \{ z  \mapsto v \mid z \mapsto u \sqcup v \in P' \}$ such that $t'\sigma_1 = t_1$ and $t' \sigma_2 = t_2$. Note that the pairs $u \sqcup v$ are not commutative.

\label{appendix:runningex}

\begin{example}
\label{ex:aunif}
Let $t_1 = (\mathit{cons}\,(\mathit{succ}\,x_1)\,(\mathit{cons}\,\mathit{zero}\,l_1))$ and $t_2 = (\mathit{cons}\,x_2\,(\mathit{cons}\,(\mathit{succ}\,x_2)\,l_2))$. Using Plotkin's algorithm on the input $\langle z, \{ z \mapsto t_1 \sqcup t_2 \} \rangle$ (note that $z$ is fresh w.r.t. $\var(t_1) \cup \var(t_2)$) we obtain:
\vspace{-1ex}
{
\begin{align*}
\langle z, \{ z \mapsto t_1 \sqcup t_2 \} \rangle  &=\\
\langle z, \{ z \mapsto (\mathit{cons}\,(\mathit{succ}\,x_1)\,(\mathit{cons}\,\mathit{zero}\,l_1)) \sqcup (\mathit{cons}\,x_2\,(\mathit{cons}\,(\mathit{succ}\,x_2)\,l_2)) \} \rangle  &\rightsquigarrow\\
\langle z[(\mathit{cons}\,z_1\, z_2)/z], \{ z_1 \mapsto (\mathit{succ}\,x_1) \sqcup x_2, z_2\mapsto (\mathit{cons}\,\mathit{zero}\,l_1) \sqcup (\mathit{cons}\,(\mathit{succ}\,x_2)\,l_2) \}  \rangle & =\displaybreak[0]\\
\langle (\mathit{cons}\,z_1\, z_2), \{ z_1 \mapsto (\mathit{succ}\,x_1) \sqcup x_2, z_2 \mapsto (\mathit{cons}\,\mathit{zero}\,l_1) \sqcup (\mathit{cons}\,(\mathit{succ}\,x_2)\,l_2) \} \rangle& \rightsquigarrow\\
\langle (\mathit{cons}\,z_1\, z_2)[(\mathit{cons}\,z_3\,z_4)/z_2], \{ z_1 {\mapsto} (\mathit{succ}\,x_1)\sqcup x_2, z_3 {\mapsto} \mathit{zero} \sqcup\! (\mathit{succ}\,x_2), z_4 {\mapsto}\, l_1 {\sqcup}\, l_2 \} \rangle& = \\
\langle (\mathit{cons}\,z_1\, (\mathit{cons}\,z_3\,z_4)), \{ z_1 \mapsto (\mathit{succ}\,x_1) \sqcup x_2, z_3 \mapsto \mathit{zero} \sqcup (\mathit{succ}\,x_2), z_4 \mapsto l_1 \sqcup l_2 \}\rangle \not\rightsquigarrow & 
\end{align*}
}

The \lgg of the term patterns $t_1$ and $t_2$ is the term pattern $t \eqbydef (\mathit{cons}\,z_1\, (\mathit{cons}\,z_3\,z_4))$ while the substitutions $\sigma_1 = \{z_1 \mapsto (\mathit{succ}\,x_1), z_3 \mapsto  \mathit{zero}, z_4 \mapsto l_1 \}$ and $\sigma_2 = \{z_1 \mapsto x_2, z_3 \mapsto (\mathit{succ}\,x_2), z_4 \mapsto l_2 \}$   satisfy $t\sigma_1  = t_1$ and $t\sigma_2 = t_2$. 
The generated variables $z_1, z_2, z_3, z_4$ occur at most once in the computed \lgg, and $\var(t) = \dom(\sigma_1) = \dom(\sigma_2)$.
\end{example}

The $\rightsquigarrow^!$ in Example~\ref{ex:aunif} means that $\rightsquigarrow$ has been applied repeatedly until $\langle t, P \rangle$ is in solved form.\\

Antiunification problems are encoded 
as \ML patterns as below:

\begin{definition}[Antiunification problem]
\label{def:auaspattern}
For each antiunification problem $\langle t, P \rangle$ we define an \ML pattern 
$$\phi^{\langle t, P\rangle} \eqbydef \exists\overline{z}. t \land \big(\phi^{\sigma_1} \vee \phi^{\sigma_2}\big),$$
\noindent
where $\sigma_1 =\{ z  \mapsto u \mid z \mapsto u \sqcup v \in P \}$, $\sigma_2  = \{ z  \mapsto v \mid z \mapsto u \sqcup v \in P \}$, and $\var(t) = \dom(\sigma_1) = \dom(\sigma_2) = \overline{z}$.
\end{definition}

\begin{example}
  \label{ex:step1}
  Here are the corresponding encodings for the intermediate antiunification problems that are generated during the execution shown in Example~\ref{ex:aunif}:
  
  {
  \begin{enumerate}
  \item 
  $\langle z, \{ z \mapsto (\mathit{cons}\,(\mathit{succ}\,x_1)\,(\mathit{cons}\,\mathit{zero}\,l_1)) \sqcup (\mathit{cons}\,x_2\,(\mathit{cons}\,(\mathit{succ}\,x_2)\,l_2)) \} \rangle$
  is encoded as\\
  \hfill $\exists z. z \land \big(z = (\mathit{cons}\,(\mathit{succ}\,x_1)\,(\mathit{cons}\,\mathit{zero}\,l_1)) \vee z =  (\mathit{cons}\,x_2\,(\mathit{cons}\,(\mathit{succ}\,x_2)\,l_2)) \big)$;
  
  \item $\langle (\mathit{cons}\,z_1\, z_2), \{ z_1 \mapsto (\mathit{succ}\,x_1) \sqcup x_2, z_2 \mapsto (\mathit{cons}\,\mathit{zero}\,l_1) \sqcup (\mathit{cons}\,(\mathit{succ}\,x_2)\,l_2) \} \rangle$ is encoded as:
  
  $\exists\{ z_1, z_2\}. (\mathit{cons}\,z_1\, z_2) \land \Big(
  \big(z_1 =  (\mathit{succ}\,x_1) \land z_2 = (\mathit{cons}\,\mathit{zero}\,l_1) \big)
  \vee 
  \big(z_1 =  x_2 \land z_2 = (\mathit{cons}\,(\mathit{succ}\,x_2)\,l_2) \big)\Big)$;
  
  \item 
  $\langle (\mathit{cons}\,z_1\, (\mathit{cons}\,z_3\,z_4)), \{ z_1 \mapsto (\mathit{succ}\,x_1) \sqcup x_2, z_3 \mapsto \mathit{zero} \sqcup (\mathit{succ}\,x_2), z_4 \mapsto l_1 \sqcup l_2 \}\rangle$
   is encoded as:

  $\exists \{z_1, z_3, z_4\}. (\mathit{cons}\,z_1\, (\mathit{cons}\,z_3\,z_4)) \land$ 
  
  \hfill $ \Big( \big(z_1 = (\mathit{succ}\,x_1) \land z_3 = \mathit{zero} \land z_4 = l_1\big) \vee \big(z_1 = x_2 \land z_3 = (\mathit{succ}\,x_2) \land z_4 = l_2\big) \Big)$.
  \end{enumerate}
  }
  \end{example}\vspace{-1ex}
  
  Note that the encodings shown in Example~\ref{ex:step1} are all equivalent. Also, remember that the scope of the quantifiers extends as much as possible to the right.

\begin{restatable}{lemma}{stepaunif}
  \label{lem:stepaunif}
  If $\langle t_i, P_i \rangle \leadsto \langle t_{i+1}, P_{i+1} \rangle$ is a step performed using Plotkin's antiunification rule, then\\ $$\mathit{TERM(S, F)} \models\phi^{\langle t_i, P_i \rangle} \leftrightarrow \phi^{\langle t_{i+1}, P_{i+1} \rangle}.$$ 
  \end{restatable}

The soundness theorem shown below is a direct consequence of Lemma~\ref{lem:stepaunif}
:\\

\vspace{-1ex}
\begin{restatable}{theorem}{antiunif}{\bf (Soundness)}
\label{th:antiunif}
Let $t_1$ and $t_2$ be two term patterns and $z$ a variable such that $z \not\in \var(t_1) \cup \var(t_2)$. If $\langle z, \{ z\mapsto t_1 \sqcup t_2 \} \rangle \rightsquigarrow^! \langle t, P \rangle$, then  $\mathit{TERM(S, F)} \models (t_1 \lor t_2) \leftrightarrow \phi^{\langle t, P \rangle}$.
\end{restatable}

The above results are proved using the semantical \ML satisfaction relation ($\models$). Recall that our goal is to generate proof objects, and thus, we want  to prove the above results using the \ML proof system. We address this challenge in the following section.

\section{Generating Proof Objects}
\label{sec:aunifgen}


Our method for generating proof objects is generic in the sense that it can be used for a larger class of term-algebra-based algorithms (e.g., unification, antiunification). 

A proof object is represented by a sequence of lines of the form:\\[2ex]
\centerline{
\begin{tabular}{|c|c|c|}
\hline
$~~~k~~~$ & ~~~derived pattern ~~~ &~~~ justification~~~\\
\hline
\end{tabular}
}\\[2ex]
where $k$ is the step index and the justification mentions the applied inference rule and the step index of the premises of the rule (if any). The step index of the premises should be smaller than $k$, i.e., the premises are justified by the previous lines.\\

Our method is sketched as follows:

%
%
\begin{enumerate}
\item We consider algorithms that transform a pattern $\varphi$ into an equivalent one $\varphi'$.
So, a proof object  $\mathit{proofObj}$ has to be generated for $\varphi\leftrightarrow \varphi'$.

\item The execution of such algorithms for an input $\varphi$ produces a sequence of intermediate patterns $\varphi_1,\ldots,\varphi_{n-1}$ such that  $\varphi^{i-1}\leftrightarrow\varphi^i$, $0 < i \le n$, where $\varphi_0 \eqbydef \varphi$ and $\varphi_n \eqbydef\varphi'$. So, a proof object for each $\varphi^{i-1}\leftrightarrow\varphi^i$ has to be generated in order to build $\mathit{proofObj}$.

\item Assuming that $\varphi^i$ is obtained from $\varphi^{i-1}$ by applying a generic step of the algorithm, 
we design a proof schema for this step s.t. the instance of this schema for  $\varphi^{i-1}$ and $\varphi^i$ produces a proof object $\mathit{proofObj}_i$ for  $\varphi^{i-1}\leftrightarrow\varphi^i$.

\item The proof object  $\mathit{proofObj}$ for the equivalence $\varphi\leftrightarrow \varphi'$ is obtained by the composition
$$\mathit{proofObj}_1;\ldots ;\mathit{proofObj}_n;\mathit{proofObj}_{n+1},$$
\noindent
where $\mathit{proofObj}_{n+1}$
connects the other proofs objects by transitivity so that the result is a proof object.
\end{enumerate}

\noindent
The approach from~\cite{fm} for unification can be formalised now as an instance of this method.
\vspace{-1ex}


\subsection{Generating proof objects for antiunification}
The goal is to generate proof objects for the equivalence $\mathit{TERM(S, F)} \models (t_1 \lor t_2) \leftrightarrow \phi^{\langle t, P \rangle}$ (cf. Theorem~\ref{th:antiunif}).
For an input antiunification problem $\langle t_0, P_0 \rangle \eqbydef \langle z , z \mapsto t_1 \sqcup t_2\rangle$, Plotkin's algorithm generates a sequence of antiunification problems until it reaches the final pair $\langle t_k, P_k \rangle \eqbydef \langle t, P \rangle$ which contains the \lgg:\\[1ex]
\centerline{$\langle t_0, P_0 \rangle \rightsquigarrow \cdots \rightsquigarrow\langle t_i, P_i \rangle \rightsquigarrow \cdots \rightsquigarrow \langle t_k, P_k \rangle.$}\\[1ex]
The key observation is that the \ML encodings of the antiunification problems from the above sequence are all equivalent (cf. Lemma~\ref{lem:stepaunif}):\\[1ex]
\centerline{$\phi^{\langle t_0, P_0 \rangle} \leftrightarrow \cdots \leftrightarrow\phi^{\langle t_i, P_i \rangle} \leftrightarrow \cdots \leftrightarrow \phi^{\langle t_k, P_k \rangle}.$}\\[1ex]
\noindent

The proof object for $t_1\lor t_2 \leftrightarrow \phi^{\langle t, P \rangle}$ is obtained by instantiating the next schema:\\[1ex]
\centerline{
  \orgen;(\decaunif)$^k$;($\eqvtranz$)$^k,$
}\\[1ex]
 \noindent
where: 
\begin{itemize}[topsep=0pt, partopsep=1pt, itemsep=1pt]
  \item $k$ is the number of applications of Plotkin's rule;
  \item \orgen is the proof schema which corresponds to the initial equivalence  $t_1 \vee t_2 \leftrightarrow \phi^{\langle t_0, P_0\rangle}$.
  Recall that $\phi^{\langle t_0, P_0\rangle} = \phi^{\langle z, z \mapsto t_1 \sqcup t_2\rangle} \eqbydef\exists z. z \land (z = t_1 \vee z = t_2)$;
  
  \item \decaunif is the proof schema corresponding to the equivalences $\phi^{\langle t_i, P_i \rangle} \leftrightarrow \phi^{\langle t_{i+1}, P_{i+1} \rangle}$, with $i \in \{0, \ldots, k-1\}$.   
All these equivalences are obtained by applying a generic step of the algorithm. 
The schema \decaunif corresponds to this generic step.
  
  \item Finally, using the transitivity of $\leftrightarrow$ 
  $k$ times we obtain a proof object for 
  $t_1 \vee t_2 \leftrightarrow \phi^{\langle t_k, P_k \rangle}$.\\
  \end{itemize}
  
The proof schema for \orgen and \decaunif are presented in Sections~\ref{sec:orgen} and~\ref{sec:decaunif}. Both use the proof rules in Figure~\ref{fig:proofsystem} and the macro rules  in Section~\ref{sec:macrorules}. 
The proof schema for \decaunif  uses two additional (sub)schema \existsgen' (shown in Section~\ref{sec:existsgenprim}) and \adec (shown in Section~\ref{sec:adec}).

Example~\ref{ex:toplevelproof} shows a high-level proof object where we apply \orgen once, \decaunif and $\eqvtranz$ twice. This is because the antiunification rule has been applied two times to obtain the \lgg. The exact (low-level) proof object corresponding to this example is obtained by instantiating the proof schemata for \orgen and \decaunif (presented later in Sections~\ref{sec:orgen} and~\ref{sec:decaunif}).

\begin{example}
  \label{ex:toplevelproof}

We show here the proof object corresponding to the execution from Example~\ref{ex:aunif}, where we use the encodings in Example~\ref{ex:step1}. The structure of the proof object follows the 
schema \orgen;(\decaunif)$^k$;($\eqvtranz$)$^k$, where $k = 2$:

  \begin{center}
  \selectfont
  \begin{tabular}{|l|l|l|}
  \hline
  $(1)$ &$  t_1 \vee t_2 
   \leftrightarrow$ & \\
   & \hfill $\exists z. z \land \big(z=(\mathit{cons}\,(\mathit{succ}\,x_1)\,(\mathit{cons}\,\mathit{zero}\,l_1)) \vee 
  z=(\mathit{cons}\,x_2\,(\mathit{cons}\,(\mathit{succ}\,x_2)\,l_2))\big)$ & \orgen\\
  \hline
  $(2.1)$ &$ \exists z. z \land$ & \\
  &\hfill $ \big(z=(\mathit{cons}\,(\mathit{succ}\,x_1)\,(\mathit{cons}\,\mathit{zero}\,l_1)) \vee 
  z=(\mathit{cons}\,x_2\,(\mathit{cons}\,(\mathit{succ}\,x_2)\,l_2))\big) 
   \leftrightarrow $ & \\
   & $\exists \{z_1,z_2\}. (\mathit{cons}\,z_1\, z_2) \land$ &\\
   & \hfill$\Big(
  \big(z_1=(\mathit{succ}\,x_1) \land z_2=(\mathit{cons}\,\mathit{zero}\,l_1) \big)
  \vee 
  \big(z_1=x_2 \land z_2=(\mathit{cons}\,(\mathit{succ}\,x_2)\,l_2) \big)\Big)$ & \decaunif\\
  \hline
  $(2.2)$ & $\exists \{z_1, z_2\}. (\mathit{cons}\,z_1\, z_2) \land$&\\ 
  &\hfill$\Big(
  \big(z_1\!=\!(\mathit{succ}\,x_1) \land z_2\!=\!(\mathit{cons}\,\mathit{zero}\,l_1) \big)
  \vee 
  \big(z_1\!=\!x_2\land z_2\!=\!(\mathit{cons}\,(\mathit{succ}\,x_2)\,l_2) \big)\Big) \leftrightarrow$ & \\
  
  & $\exists \{z_1, z_3, z_4\}. (\mathit{cons}\,z_1\, (\mathit{cons}\,z_3\,z_4)) \land $ &\\ 
  & \hfill$\Big( \big(z_1 = (\mathit{succ}\,x_1) \land z_3 = \mathit{zero} \land z_4 = l_1\big) \vee \big(z_1 = x_2 \land z_3 = (\mathit{succ}\,x_2) \land z_4 = l_2\big) \Big)$ & \decaunif\\
  \hline
  $(3.1)$ & $t_1 \vee t_2  \leftrightarrow$ & \\
  & $\exists \{z_1, z_2\}. (\mathit{cons}\,z_1\, z_2) \land$& $\eqvtranz$:\\ 
  &\hfill$\Big(
  \big(z_1 =  (\mathit{succ}\,x_1) \land z_2 = (\mathit{cons}\,\mathit{zero}\,l_1) \big)
  \vee 
  \big(z_1 =  x_2 \land z_2 = (\mathit{cons}\,(\mathit{succ}\,x_2)\,l_2) \big)\Big)$ & $1, 2.1$ \\
  \hline
  $(3.2)$ & $t_1 \vee t_2  \leftrightarrow$ & $\eqvtranz$:\\
  & $\exists \{z_1, z_3, z_4\}. (\mathit{cons}\,z_1\, (\mathit{cons}\,z_3\,z_4)) \land $ & $3.1, 2.2$ \\ 
  & \hfill $\Big( \big(z_1 = (\mathit{succ}\,x_1) \land z_3 = \mathit{zero} \land z_4 = l_1\big) \vee \big(z_1 = x_2 \land z_3 = (\mathit{succ}\,x_2) \land z_4 = l_2\big) \Big)$ &  \\
  \hline
  \end{tabular}
  
  \end{center}

\end{example}


\bigskip  
Generating proof objects for antiunification turns out to be more complex than in the unification case from~\cite{fm}.
Plotkin's algorithm generates fresh variables at each step. 
These variables are existentially quantified in the corresponding \ML encodings and handling these quantifiers in proofs is difficult.
The main difficulty comes from the fact that most of the times the application of a \ML proof rule requires a lot of preparation work:
you have to isolate the goal that can be proved using a particular \ML proof rule, and then find a way to put back the existential quantifiers.
Also, you have to make sure that the proof objects generated by \decaunif remain composable, so that $\eqvtranz$ can be applied. 
This is why we use several macro rules in addition to the \ML proof system.
\vspace{-1ex}


\subsection{The Macro Rules}
\label{sec:macrorules}


Our approach uses the additional rules shown in Figure~\ref{fig:macro-rules}\footnote{Proving these rules using the proof system in Figure~\ref{fig:proofsystem} is out of scope of this paper.}.
The first part includes three macro rules. 
$\econtext$ enables the replacement of a formula with an equivalent one under the $\exists$ quantifier. 
$\escope$ extends the scope of $\exists$ over formulas that do not contain variables that can be captured. 
$\ecollapse$  is useful when existentially quantified formulas can be collapsed under a single quantifier.

The second part includes two macro rules which are consequences of the specification $\mathit{TERM(S, F)}$ (cf. Proposition~\ref{prop:rulestermalg}).
\substexists states that $t \land (z = u)$ is equivalent to $t[u/z]$ under the existential quantifier which binds $z$. 
\existsgen allows one to replace subterms $\overline{t} \eqbydef t_1 \ldots t_n$ of a term with existentially quantified fresh variables $\overline{y} \eqbydef y_1 \ldots y_n$. To obtain an equivalent formula, the constraints $\overline{y} = \overline{t}$ 
are added.

\begin{figure}[h]
  \centering
  \renewcommand{\arraystretch}{1.25}
  \vspace{-2ex}
  \begin{tabular}{|lrl|}
  \hline
  \multicolumn{3}{|c|}{\bf Additional proof rules}\\ 
  \hline
  \econtext && \infer{\big(\exists \overline{x}.\varphi_1 \land \varphi_2\big) \leftrightarrow \exists \overline{x}.(\varphi_1 \land \varphi'_2)}{\varphi_2 \leftrightarrow \varphi'_2} \\
  \escope && $\big(\big(\exists \overline{x}. \varphi_1\big) \odot \varphi_2\big) \leftrightarrow \exists \overline{x}. (\varphi_1 \odot \varphi_2), \textrm{if~}\overline{x} \not\in \mathit{free}(\varphi_2)$\\
  \ecollapse && $\big((\exists \overline{x}. \varphi_1) \vee (\exists \overline{x}. \varphi_2)\big) \leftrightarrow \exists \overline{x}. (\varphi_1 \vee \varphi_2)$\\
  \hline
  
  \multicolumn{3}{|c|}{\bf Term algebra specific proof rules}\\ 
  \hline
  $\substexists$ && $\exists z. t \land (z = u) \leftrightarrow t[u/z], \textrm{if~}z \not\in\var(u)$\\
  
  $\existsgen$ && $z=(f\, \overline{t}) \leftrightarrow \exists \overline{y}. z=(f\, \overline{y}) \wedge \overline{y} = \overline{t}$, $\textrm{if~}\overline{y} \not\in \var\big((f\, \overline{t})\big) \cup \{z\}$ \\ 
  \hline
  \end{tabular}
  \caption{The macro rules used to generate proof objects for antiunification. The $\odot$ is a placeholder for one logical operator in the set $\{\land, \leftrightarrow\}$. Also, unless explicitly delimited using parentheses, the scope of the quantifiers extends as much as possible to the right.}
  \label{fig:macro-rules}
  \end{figure}

\subsection{Proof object schema \orgen}
\label{sec:orgen}

The first step of our method is to establish the equivalence between the disjunction $t_1 \vee t_2$  and the encoding of the initial unification problem $\langle z , z \mapsto t_1 \sqcup t_2\rangle$, that is  $\big(\exists z. z \land (z = t_1)\big) \vee  \big(\exists z.   z \land (z = t_2)\big)$. 
This is done via the proof object schema \orgen, which is shown below.

To shorten our presentation, the \textsc{ModusPonens} rule from Figure~\ref{fig:proofsystem} is applied here directly over a double implication $\leftrightarrow$ (instead of $\rightarrow$). 
For the steps k$\mathop{+}$6 and k$\mathop{+}$8 we use \textsc{Propositional} to justify two trivial equivalences: the first says that $\phi_1 \vee \phi_2 \leftrightarrow \phi_1' \vee \phi_2'$ if $\phi_1 \leftrightarrow \phi_1'$ and $\phi_2 \leftrightarrow \phi_2'$; the second is just a well-known distributivity property.\\
\begin{center}
\begin{tabular}{|l|l|l|}
\hline
(k) &$\big(\exists z. t_1 \leftrightarrow  z \land (z = t_1)\big)$ & \substexists (note: $z[t_1/z] = t_1$) \\
\hline
(k$\mathop{+}$1) &$\big(\exists z. t_1 \leftrightarrow  z \land (z = t_1)\big) \leftrightarrow \big(t_1 \leftrightarrow \exists z. z \land (z = t_1)\big) $ & \escope \\
\hline
(k$\mathop{+}$2) & $t_1 \leftrightarrow \exists z. z \land (z = t_1)$ & \textsc{ModusPonens}: k, k+1\\
\hline
(k$\mathop{+}$3) &$\big(\exists z. t_2 \leftrightarrow  z \land (z = t_2)\big)$ & \substexists (note: $z[t_2/z] = t_2$) \\
\hline
(k$\mathop{+}$4) &$\big(\exists z. t_2 \leftrightarrow  z \land (z = t_2)\big) \leftrightarrow \big(t_2 \leftrightarrow \exists z. z \land (z = t_2)\big)$ & \escope \\
\hline
(k$\mathop{+}$5) & $t_2 \leftrightarrow \exists z. z \land (z = t_2)$ & \textsc{ModusPonens}: k+3, k+4\\
\hline

(k$\mathop{+}$6) & $t_1 \vee t_2 \leftrightarrow \big(\exists z. z \land (z = t_1)\big) \vee  \big(\exists z.   z \land (z = t_2)\big)$ & \textsc{Propositional}: k+2, k+5\\
%
\hline

(k$\mathop{+}$7) & $\Big(\big(\exists z.  z \land (z = t_1)\big) \vee  \big(\exists z. z \land (z = t_2)\big) \Big)\leftrightarrow$ &\\
&  \hfill$\exists z. \big(z \land (z = t_1)\big) \vee  \big(  z \land (z = t_2)\big)$ & \ecollapse \\
\hline

(k$\mathop{+}$8) & $\big( z \land (z = t_1)\big) \vee  \big(z \land (z = t_2)\big) \leftrightarrow $ &  \\
& $ z \land \big((z = t_1) \vee  (z = t_2)\big)$ & \textsc{Propositional} \\
\hline

(k$\mathop{+}$9) & $\exists z. \big( z \land (z = t_1)\big) \vee  \big(z \land (z = t_2)\big) \leftrightarrow$ &  \\
& $ \exists z. z \land \big((z = t_1) \vee  (z = t_2)\big)$ & \econtext: k$\mathop{+}$8 \\
\hline

(k$\mathop{+}$10) & $t_1 \vee t_2 \leftrightarrow \exists z. \big(z \land (z = t_1)\big) \vee  \big(  z \land (z = t_2)\big)$ & $\eqvtranz$: k$\mathop{+}$6, k$\mathop{+}$7 \\
\hline

(k$\mathop{+}$11) & $t_1 \vee t_2 \leftrightarrow \exists z. z \land \big((z = t_1) \vee  (z = t_2)\big)$ & $\eqvtranz$: k$\mathop{+}$10, k$\mathop{+}$9\\

\hline
\end{tabular}
\end{center}

\aaa{Am mai adaugat niste propozitii la fiecare dintre sectiunile care urmeaza: pentru e-gen', dec si step. }

\subsection{Proof schema \decaunif}
\label{sec:decaunif}

The proof schema of \decaunif used two other (sub)schemata \existsgen' and \adec. We explain these first, and them we present the proof schema for \decaunif.

\subsubsection{Proof object schema \existsgen'}
\label{sec:existsgenprim}

Recall that $\existsgen$ (Figure~\ref{fig:macro-rules} -- term algebra specific rule) establishes an equivalence between
$z=(f\, \overline{t})$ and $\exists \overline{y}. z=(f\, \overline{y}) \wedge \overline{y} = \overline{t}$, $\textrm{if~}\overline{y} \not\in \var\big((f\, \overline{t})\big) \cup \{z\}$, which basically describes a generalisation of $(f\, \overline{t})$. 
However, most of the times $\existsgen$ is applied under a conjunction.
The proof schema $\existsgen'$ generalises the $\existsgen$ macro rule as follows:\\[1ex]
\centerline{$\big(\varphi \land z = (f\,\overline{t})\big) \leftrightarrow \exists \overline{z} . \varphi \land z = (f\,\overline{z}) \land \overline{z} = \overline{t},$}\\[1ex]
\noindent
where $(f\,\overline{t})$ denotes $(f\,t_1\,\ldots\,t_n)$, $ (f\,\overline{z})$ stands for  $(f\,z_1\,\ldots\,z_n)$, and $\overline{z} = \overline{t}$ denotes the conjunction $\bigwedge_{i=1}^n z_i = t_i$. Note that $\varphi$ is safely introduced under the existential quantifier. 
The proof schema \existsgen' is shown below:

\begin{center}
\begin{tabular}{|l|l|l|}
\hline
(k) &$z = (f\,\overline{t}) \leftrightarrow \exists \overline{z} . z = (f\,\overline{y}) \land \overline{y} = \overline{t}$ & $\existsgen$ \\
\hline
(k$\mathop{+}$1) &$\big(\varphi \land z = (f\,\overline{t})\big) \leftrightarrow $ & \\
& \hfill $ \varphi \land \exists \overline{z} . z = (f\,\overline{z}) \land \overline{z} = \overline{t}$ & \textsc{Propositional}:k \\
\hline
(k$\mathop{+}$2) &$\big(\varphi \land \exists \overline{z} . z = (f\,\overline{z}) \land \overline{z} = \overline{t}\big) \leftrightarrow $ & \\
& \hfill $ \exists  \overline{z} .  \varphi \land  z = (f\,\overline{z}) \land \overline{z} = \overline{t}$ & $\escope$, $\var(\varphi)\cap\{z_1, \ldots, z_n\} = \emptyset$ \\
\hline
(k$\mathop{+}$3) &$\big(\varphi \land z = (f\,\overline{t})\big) \leftrightarrow $ & \\
& \hfill $ \exists \overline{z} . \varphi \land  z = (f\,\overline{z}) \land \overline{z} = \overline{t}$ & $\eqvtranz$: k$\mathop{+}$1, k$\mathop{+}$2\\
\hline
\end{tabular}
\end{center}

\noindent
At step $k\mathop{+}1$ we use \textsc{Propositional}, in particular, we use this property: if $\varphi_1 \leftrightarrow \varphi_2$ then $\varphi \land \varphi_1 \leftrightarrow \varphi \land \varphi_2$.
This schema is applied in a certain context, where $\var(\varphi)\cap\{z_1, \ldots, z_n\} = \emptyset$, because $z_1, \ldots, z_n$ are always fresh variables introduced by Plotkin's antiunification algorithm.

\subsubsection{Proof schema {\adec}}
\label{sec:adec}
Once we have equivalent forms for $z=(f\, \overline{t})$ (i.e., $\exists \overline{y}. z=(f\, \overline{y}) \wedge \overline{y} = \overline{t}$, cf. $\existsgen'$), we are now ready to tackle disjunctions $(f\,\overline{u}) \vee (f\,\overline{v})$.
\adec captures a decomposition: $(f\,\overline{u}) \vee (f\,\overline{v})$ is equivalent to a conjunction between $(f\,\overline{z})$ and $(\overline{z} = \overline{u}) \vee (\overline{z} = \overline{v})$, where again $\overline{z} = \{ z_1\ldots\,z_n \}$ are existentially quantified. In addition, \adec performs the decomposition under a conjunction:\\[1ex]
\centerline{$
\big((\varphi \land z = (f\,\overline{u})) \vee (\varphi' \land z = (f\,\overline{v})) \big) \leftrightarrow \exists \overline{z} . z = (f\,\overline{z}) \land \big((\varphi \land \overline{z} = \overline{u}) \vee  (\varphi' \land \overline{z} = \overline{v})\big),
$}\\[1ex]
\noindent
where $(f\,\overline{u})$ means $(f\,u_1\,\ldots\,u_n)$, $ (f\,\overline{z})$ stands for  $(f\,z_1\,\ldots\,z_n)$, and the equality $\overline{z} = \overline{u}$ denotes $\bigwedge_{i=1}^n z_i = u_i$. 
Similarly, $(f\,\overline{v})$ denotes $(f\,v_1\,\ldots\,v_n)$ and $\overline{z} = \overline{v}$ is $\bigwedge_{i=1}^n z_i = v_i$. 
The schema for \adec uses \existsgen':\\

\begin{center}
\begin{tabular}{|l|l|l|}
\hline
(k) & $\big(\varphi \land z\mathop{=}(f\,\overline{u})\big) \leftrightarrow \exists \overline{z} . \varphi \land  z\mathop{=}(f\,\overline{z}) \land \overline{z}\mathop{=}\overline{u}$ & \existsgen'\\ \hline

(k$\mathop{+}$1) & $\big(\varphi' \land z\mathop{=}(f\,\overline{v})\big) \leftrightarrow \exists \overline{z} . \varphi' \land  z\mathop{=}(f\,\overline{z}) \land \overline{z}\mathop{=}\overline{v}$ & \existsgen'\\ \hline

(k$\mathop{+}$2) & $\big(\varphi \land z\mathop{=}(f\,\overline{u})\big) \lor  \big(\varphi' \land z\mathop{=}(f\,\overline{v})\big) \leftrightarrow $ &  \textsc{Propositional}:\\
& \hfill $\big(\exists \overline{z} . \varphi \land  z\mathop{=}(f\,\overline{z}) \land \overline{z}\mathop{=}\overline{u}\big) \vee 
\exists \overline{z} . \varphi' \land  z\mathop{=}(f\,\overline{z}) \land \overline{z}\mathop{=}\overline{v}$ &  k$\mathop{+}$1, k$\mathop{+}$2\\ \hline

(k$\mathop{+}$3) & $\big(\exists \overline{z} . \varphi \land  z\mathop{=}(f\,\overline{z}) \land \overline{z}\mathop{=}\overline{u}\big) \vee 
\exists \overline{z} . \varphi' \land  z\mathop{=}(f\,\overline{z}) \land \overline{z}\mathop{=}\overline{v} \leftrightarrow $ & \\
& \hfill $\exists \overline{z} . \big( \varphi \land  z\mathop{=}(f\,\overline{z}) \land \overline{z}\mathop{=}\overline{u}\big) \vee 
 \varphi' \land  z\mathop{=}(f\,\overline{z}) \land \overline{z}\mathop{=}\overline{v}$ & \ecollapse \\ \hline

(k$\mathop{+}$4) & $\big(\varphi \land z\mathop{=}(f\,\overline{u})\big) \lor  \big(\varphi' \land z\mathop{=}(f\,\overline{v})\big) \leftrightarrow $ & \\
& \hfill $\exists \overline{z} . \big( \varphi \land  z\mathop{=}(f\,\overline{z}) \land \overline{z}\mathop{=}\overline{u}\big) \vee 
 \varphi' \land  z\mathop{=}(f\,\overline{z}) \land \overline{z}\mathop{=}\overline{v}$ & $\eqvtranz$: k$\mathop{+}$2, k$\mathop{+}$3 \\ \hline

(k$\mathop{+}$5) & $\big( \varphi \land  z\mathop{=}(f\,\overline{z}) \land \overline{z}\mathop{=}\overline{u}\big) \vee 
 \varphi' \land  z\mathop{=}(f\,\overline{z}) \land \overline{z}\mathop{=}\overline{v} \leftrightarrow$ & \\
 & \hfill $z\mathop{=}(f\,\overline{z}) \land \big( (\varphi \land  \overline{z}\mathop{=}\overline{u}) \vee 
 (\varphi'  \land \overline{z}\mathop{=}\overline{v})\big)$ & \textsc{Propositional} \\ \hline

(k$\mathop{+}$6) & $\exists \overline{z}. \big( \varphi \land  z\mathop{=}(f\,\overline{z}) \land \overline{z}\mathop{=}\overline{u}\big) \vee 
 \varphi' \land  z\mathop{=}(f\,\overline{z}) \land \overline{z}\mathop{=}\overline{v} \leftrightarrow$ & \\
 & \hfill $\exists \overline{z}. z\mathop{=}(f\,\overline{z}) \land \big( (\varphi \land  \overline{z}\mathop{=}\overline{u}) \vee 
 (\varphi'  \land \overline{z}\mathop{=}\overline{v})\big)$ & \econtext: k$\mathop{+}$5 \\ \hline

(k$\mathop{+}$7) & $\big(\varphi \land z\mathop{=}(f\,\overline{u})\big) \lor  \big(\varphi' \land z\mathop{=}(f\,\overline{v})\big) \leftrightarrow $ & \\
& \hfill $\exists \overline{z}. z\mathop{=}(f\,\overline{z}) \land \big( (\varphi \land  \overline{z}\mathop{=}\overline{u}) \vee 
 (\varphi'  \land \overline{z}\mathop{=}\overline{v})\big)$ & $\eqvtranz$: k$\mathop{+}$4, k$\mathop{+}$6 \\ 
\hline
\end{tabular}
\end{center}

\subsubsection{Proof schema \decaunif}
\label{sec:decaunif}
Recall that in each step $\langle t_i,P_i \rangle \rightsquigarrow \langle t_{i+1}, P_{i+1} \rangle$, $t_{i+1}$ is a generalisation of $t_{i}$, and both $t_{i}$ and $t_{i+1}$ are generalisations of the initial term patterns.
Also, recall that both $\phi^{\langle t_i,P_i \rangle}$ and $\phi^{\langle t_{i+1},P_{i+1} \rangle}$ are existentially quantified conjunctions between a term pattern (e.g., the generalisations $t_i$, $t_{i+1}$) and a predicate (cf. Definition~\ref{def:auaspattern}).
Our final step is to add the missing existential quantifiers and the missing term patterns (generalisations) to the equivalences obtained using \adec. 
The schema which does all the above is summarised below:\\
\begin{center}
\begin{tabular}{|l|l|l|}
\hline
(k$\mathop{+}$1) &  $\big(\varphi \land z\mathop{=}(\mathit{f}\,\overline{u})\big) \vee \big(\varphi' \land z\mathop{=}(\mathit{f}\,\overline{v})\big) \leftrightarrow $ & \\
& \hfill $ \exists \overline{z}. z\mathop{=}(f\,\overline{z}) \land \big( (\varphi \land  \overline{z}\mathop{=}\overline{u}) \vee 
 (\varphi'  \land \overline{z}\mathop{=}\overline{v})\big)$ &  \adec \\

\hline

(k$\mathop{+}$2) & $\exists\overline{x}.t \land \big(\varphi \land z\mathop{=}(\mathit{f}\,\overline{u})\big) \vee \big(\varphi' \land z\mathop{=}(\mathit{f}\,\overline{v})\big) \leftrightarrow$ &  $z \in \overline{x}\mathop{=}\var(t)$ \\
& \hfill $\exists\overline{x}.t \land \exists \overline{z}. z\mathop{=}(f\,\overline{z}) \land \big( (\varphi \land  \overline{z}\mathop{=}\overline{u}) \vee 
 (\varphi'  \land \overline{z}\mathop{=}\overline{v})\big)$ & \econtext: k+1 \\

\hline
(k$\mathop{+}$3) & $ t \land \exists \overline{z}. z\mathop{=}(f\,\overline{z}) \land \big( (\varphi \land  \overline{z}\mathop{=}\overline{u}) \vee 
 (\varphi'  \land \overline{z}\mathop{=}\overline{v})\big) \leftrightarrow$ & $\overline{z}$  fresh  \\
& \hfill $ \exists \overline{z}. t \land z\mathop{=}(f\,\overline{z}) \land \big( (\varphi \land  \overline{z}\mathop{=}\overline{u}) \vee 
 (\varphi'  \land \overline{z}\mathop{=}\overline{v})\big)$ & \escope \\

\hline
(k$\mathop{+}$4) & $ \exists \overline{x}. t \land \exists \overline{z}. z\mathop{=}(f\,\overline{z}) \land \big( (\varphi \land  \overline{z}\mathop{=}\overline{u}) \vee 
 (\varphi'  \land \overline{z}\mathop{=}\overline{v})\big) \leftrightarrow$ & $\overline{x}\in \var(t)$  \\
& \hfill $ \exists \overline{x}. \exists \overline{z}. t \land z\mathop{=}(f\,\overline{z}) \land \big( (\varphi \land  \overline{z}\mathop{=}\overline{u}) \vee 
 (\varphi'  \land \overline{z}\mathop{=}\overline{v})\big)$ &  \econtext: k+3\\

\hline
(k$\mathop{+}$5) & $\exists\overline{x}.t \land \big(\varphi \land z\mathop{=}(\mathit{f}\,\overline{u})\big) \vee \big(\varphi' \land z\mathop{=}(\mathit{f}\,\overline{v})\big) \leftrightarrow$ &  $\eqvtranz$:\\
& \hfill $ \exists \overline{x}. \exists \overline{z}. t \land z\mathop{=}(f\,\overline{z}) \land \big( (\varphi \land  \overline{z}\mathop{=}\overline{u}) \vee 
 (\varphi'  \land \overline{z}\mathop{=}\overline{v})\big)$ &   k+2, k+4 \\

\hline

(k$\mathop{+}$6) & $\exists z.t \land z\mathop{=}(\mathit{f}\,\overline{z}) \land  \big((\varphi \land \overline{z}\mathop{=}\overline{u}) \vee (\varphi' \land \overline{z}\mathop{=}\overline{v})\big) \leftrightarrow$ &  \\
& \hfill $\big(\exists z.t \land z\mathop{=}(\mathit{f}\,\overline{z})\big) \land   \big((\varphi \land \overline{z}\mathop{=}\overline{u}) \vee (\varphi' \land \overline{z}\mathop{=}\overline{v})\big) $ &\escope \\

\hline

(k$\mathop{+}$7) & $\exists \{\overline{x}, \overline{z}\}.t \land z\mathop{=}(\mathit{f}\,\overline{z}) \land  \big((\varphi \land \overline{z}\mathop{=}\overline{u}) \vee (\varphi' \land \overline{z}\mathop{=}\overline{v})\big) \leftrightarrow$ &  \\
& \hfill $\exists \{\overline{x}, \overline{z}\}\setminus\{z\}.\big(\exists z.t \land z\mathop{=}(\mathit{f}\,\overline{z})\big) \land   \big((\varphi \land \overline{z}\mathop{=}\overline{u}) \vee (\varphi' \land \overline{z}\mathop{=}\overline{v})\big) $ &\econtext: k+6 \\

\hline
(k$\mathop{+}$8) & $\exists\overline{x}.t \land \big(\varphi \land z\mathop{=}(\mathit{f}\,\overline{u})\big) \vee \big(\varphi' \land z\mathop{=}(\mathit{f}\,\overline{v})\big) \leftrightarrow$ &  $\eqvtranz$: \\
& \hfill $\exists \{\overline{x}, \overline{z}\}\setminus\{z\}.\big(\exists z.t \land z\mathop{=}(\mathit{f}\,\overline{z})\big) \land   \big((\varphi \land \overline{z}\mathop{=}\overline{u}) \vee (\varphi' \land \overline{z}\mathop{=}\overline{v})\big) $ & k+5, k+7 \\

\hline
(k$\mathop{+}$9) & $\exists z.t \land z\mathop{=}(\mathit{f}\,\overline{z}) \leftrightarrow t[(\mathit{f}\,\overline{z})/z]$ & \substexists \\
\hline
(k$\mathop{+}$10) & $\big(\exists z.t \land z\mathop{=}(\mathit{f}\,\overline{z}) \leftrightarrow t[(\mathit{f}\,\overline{z})/z]\big) \leftrightarrow  \big(\exists z.t \land z\mathop{=}(\mathit{f}\,\overline{z})\big) \leftrightarrow t[(\mathit{f}\,\overline{z})/z] $ &  \escope\\

\hline
&&  \textsc{ModusPon.}:\\
(k$\mathop{+}$11) & $\big(\exists z.t \land z\mathop{=}(\mathit{f}\,\overline{z})\big) \leftrightarrow t[(\mathit{f}\,\overline{z})/z]$ &  k+9, k+10\\

\hline
(k$\mathop{+}$12) & $\big(\exists \{\overline{x}, \overline{z}\}.t \land z\mathop{=}(\mathit{f}\,\overline{z})\big) \land \big(\varphi \land z\mathop{=}(\mathit{f}\,\overline{u})\big) \vee \big(\varphi' \land z\mathop{=}(\mathit{f}\,\overline{v})\big)  \leftrightarrow $ & \\
&\hfill  $ \exists \{\overline{x}, \overline{z}\}\setminus\{z\}. t[(\mathit{f}\,\overline{z})/z] \land \big(\varphi \land z\mathop{=}(\mathit{f}\,\overline{u})\big) \vee \big(\varphi' \land z\mathop{=}(\mathit{f}\,\overline{v})\big) $ &  \econtext: k+11\\

\hline
(k$\mathop{+}$13) & $\exists\overline{x}.t \land \big(\varphi \land z\mathop{=}(\mathit{f}\,\overline{u})\big) \vee \big(\varphi' \land z\mathop{=}(\mathit{f}\,\overline{v})\big) \leftrightarrow$ &  $\eqvtranz$:\\

&\hfill  $ \exists \{\overline{x}, \overline{z}\}\setminus\{z\}. t[(\mathit{f}\,\overline{z})/z] \land \big(\varphi \land z\mathop{=}(\mathit{f}\,\overline{u})\big) \vee \big(\varphi' \land z\mathop{=}(\mathit{f}\,\overline{v})\big) $ &   k+8, k+12 \\
\hline

\end{tabular}
\end{center}

\medskip
\noindent
This schema uses the fact that Plotkin's algorithm generates fresh variables at each antiunification step. Here, $\overline{x} = \{x_1, \ldots, x_n\}=\var(t)$ and $\exists \{\overline{x}, \overline{z}\}$ stands for $\exists \{x_1, \ldots, x_n, z_1, \ldots, z_m\}$. At k$\mathop{+}$7, we directly use $\exists \{\overline{x}, \overline{z}\}$ instead of $\exists \{\overline{x}, \overline{z}\}\setminus\{z\}.\exists z$. 

\section{A tool for certifying antiunification}
\label{sec:prototype}
We implement a prototype~\cite{maude-tool-for-certif-aintiunif} for our proof object generation mechanism and a checker for the generated proof objects. 
The proof generator and the proof checker are implemented in Maude~\cite{allAboutMaude}.
Both tools can be used directly in Maude, but we also created a Python interface in order to facilitate the interaction of the user with the Maude tools.

The Python script takes easy-to-write specifications as inputs, it automatically calls the Maude proof generator and the proof checker behind the scenes, and outputs a proof and a checking status.

The specifications are minimally specified in an input file whose content is self-explanatory.
For example, the antiunification problem from Example~\ref{ex:aunif} is specified as:

{\fontsize{8}{10}\selectfont
\begin{lstlisting}
variables: x1, x2, l1, l2
symbols: cons, succ, zero
problem: cons(succ(x1),cons(zero,l1))=?cons(x2,cons(succ(x2),l2))
\end{lstlisting}
}

The Python script parses the input and extracts the variables, the symbols, and the antiunification problem. 
It infers automatically the arities of the symbols and throws errors when the inputs are not well-formed or arities are inconsistent (e.g., same symbol used with different number of arguments). 
Then it calls the Maude proof generator and checker in the background.
The output from Maude is postprocessed in Python and the user can inspect the proof and the checking status ({\tt true} or {\tt false}):

%
{\fontsize{8}{10}\selectfont
\begin{lstlisting}[language=c++]
> python3 ml-antiunify.py tests/samples/13_paper_cons_succ.in
Proof of: // goal: ...
// generated proof...
Checked:   true
\end{lstlisting}
}

\noindent
For this particular example, the generated proof has 84 proof lines 
as shown in the first line of Table~\ref{tbl:results}. 

We tested our prototype on larger inputs as well. Our goal was to see if the size of the generated proof objects for real-life language configurations is indeed manageable (e.g., the size should increase linearly, not exponentially). Also, we wanted to check whether our proof-generation schemas are correctly instantiated and they compose as expected to obtain the final proof objects.
The inputs that we use are inspired from the K definitions of several languages, including the K definitions of C~\cite{ellison-rosu-2012-popl} and Java~\cite{DBLP:conf/popl/BogdanasR15}. In these K definitions, the configurations  are quite big (i.e., $\sim$130 nodes for C, $\sim$65 nodes for Java). 
From these definitions we extracted some large term patterns which we use as inputs for antiunification.

Table~\ref{tbl:results} shows some of the results that we obtained\footnote{Details can be found here: \url{https://github.com/andreiarusoaie/certifying-unification-in-aml/tree/master/tests/samples\#readme}}.
The input terms in our specifications represent C and Java symbolic configurations converted into our input format. We use the specification file size to measure the input size and we use the number of proof lines to measure the size of proof object. 
As expected, the size of the generated proof objects depends on the input size. For language definitions that have larger configurations the generated proofs are big (e.g., ~5000 lines for C, and 2300 lines for Java). 

\begin{table}
  \caption{The results obtained when generating proof objects for large inputs inspired from K definitions of real-life languages (C and Java).\\[1ex]}
  \centering
    \begin{tabular}{|l|l|c|c|}
  \hline
      Language & File & File size & proof object size \\ 
       & name  & (kb) & (no. of lines) \\ 
  \hline
    list of nats	& \texttt{13\_paper\_cons\_succ.in} & 0.122&  84\\
  C& \texttt{18\_c\_declare\_local.in}& 14& 5052\\
  Java&\texttt{19\_java\_method\_invoke.in}& 6& 2352\\
  \hline
    \end{tabular}
  \label{tbl:results}
  \end{table}

\dl{cred ca putem fi mai precisi aici: dimensiunea instantelor este acceeasi cu cea a schemelor, asa ca marimea proof objectului este proportionala cu cea executiei algoritmului de unificare, care in cazul cel mai nefavorabil este dimensiunea celui mai mic termen (e.g., t2 instanta a lui t1, atunci o ss fie marimea lui t1).}

\aaa{Am adaugat cele 2 paragrafe de mai jos.}
For each step of the antiunification algorithm, the proof generator (which implements the schemas discussed in Section~\ref{sec:aunifgen}) produces a fixed number of proof lines. Therefore, the size of the proof object is directly proportional with the number of execution steps of the antiunification algorithm. In the worst-case scenario, when executed on an input $\langle z , z \mapsto t_1 \sqcup t_2\rangle$ the number of execution steps of the antiunification algorithm is given by the size of the smallest term pattern between $t_1$ and $t_2$. 
An example of worst-case scenario is when one term is an instance of the other, i.e., the \lgg $t_1$ and $t_2$ is $t_i$ with $i \in \{1,2\}$.\\

The tests that we performed on large inputs show that our proof object generation method can be implemented and used in practice. Our prototype is able to produce a correct output
(i.e., the proof objects are successfully checked by the checker) and thus, it shows that there are no corner cases that we missed in our approach.
\vspace{-1ex}


%


\section{Conclusions}
\label{sec:conclusions}
\vspace{-1ex}

In order to obtain a certified symbolic execution,
 the parameters of an execution step have to carry the proof object of their actions. Two examples of such parameters are unification and antiunification algorithms. 
\aaa{Partea pana la future work a fost modificata.}
In this paper we proposed a generic method for generating proof objects for the tasks that the K tool performs. We showed how this method works for the case of antiunification by providing schemas for generating proof objects.  
More precisely, we used Plotkin's antiunification  to normalise disjunctions and to generate the corresponding proof objects. We also provided a prototype implementation of our proof object generation technique and a checker for the generated proof objects.

The prototype generates proof objects whose size depends on the number of steps performed by the antiunification algorithm. However, the number of steps is limited by the size of the inputs. 
We successfully used our prototype on complex inputs inspired from the K semantics of C and Java. 
This indicates that our approach is practical even for large inputs.\\

\textbf{Future work}
In order to obtain a proof object that uses only the Hilbert-style proof rules, the next step is to find proof schemata for the macro rules in Figure~\ref{fig:macro-rules}. This will allow us to use the newest proof checker for \ML implemented in Metamath~\cite{checkermm}.
Both checkers (the existing Metamath and our Maude implementations) have the same functionality w.r.t. \ML proof system. 
So far, we preferred our Maude implementation because it was easier to handle macro rules. Also, we reused only the \ML syntax module written in Maude for both the checker and the proof generator.
However, the short term goal is to use the Metamath checker so that the proof object generator and the checker are completely independent.

\bibliographystyle{eptcs}
\bibliography{refs}

\newpage

\end{document}